\documentclass[twocolumn]{revtex4}

\usepackage{graphics}

\usepackage{amsfonts}
\usepackage{amsbsy}
\usepackage{amssymb}
\usepackage{amsmath}
\usepackage{graphicx}
\usepackage{color}

\newcommand\pa{\partial}

\newcommand{\beque}{\begin{equation*}}
\newcommand{\eeq}{\end{equation}}
\newcommand{\beq}{\begin{equation}}
\newcommand{\eeque}{\end{equation*}}
\newcommand{\beqnl}{\begin{eqnarray}}
\newcommand{\eeqna}{\end{eqnarray*}}
\newcommand{\beqna}{\begin{eqnarray*}}
\newcommand{\eeqnl}{\end{eqnarray}}

\newcommand\ham{{\mathcal H}}

\begin{document}
\title{Perturbative photon production in a dispersive medium.}
\author{F.~Belgiorno,$^1$ S.L.~Cacciatori,$^{2,3}$ F.~Dalla~Piazza,$^4$}

\address{$^1$ Dipartimento di Matematica, Politecnico di Milano, Piazza Leonardo 32,20133 Milano, Italy, francesco.belgiorno@polimi.it\\
$^2$ Department of Physics and Mathematics, Universit\`a dell'Insubria, Via Valleggio 11, 22100 Como, Italy, sergio.cacciatori@uninsubria.it\\
$^3$ INFN sezione di Milano, via Celoria 16, 20133 Milano, Italy\\
$^4$  Universit\`a ``La Sapienza'', Dipartimento di Matematica, Piazzale A. Moro 2, I-00185, Roma, Italy, dallapiazza@mat.uniroma1.it, f.dallapiazza@gmail.com}
%

%
\begin{abstract}
We investigate photon pair-creation in a dispersive dielectric medium induced by the presence of a spacetime 
varying dielectric constant. Our aim is to examine the possibility to observe new phenomena of pair creation 
induced by travelling dielectric perturbations e.g. created by laser pulses by means of the Kerr effect. 
In this perspective, we adopt a semi-phenomenological version of the Hopfield model in which a space-time 
dependent dielectric susceptibility appears. We focus our 
attention on perturbation theory, and provide general formulas for the photon production induced by a local 
but arbitrarily spacetime dependent refractive index perturbation. As an example, we further explore the case 
of an uniformly travelling perturbation, and provide examples of purely time-dependent 
perturbations. 
\end{abstract}
\maketitle
\section{Introduction}

Pair creation by an external field or by moving boundaries is a very interesting research field which has been explored since the birth of 
modern quantum field theory \cite{heisenberg,schwinger}. We focus on photon pair creation associated with variations of the dielectric 
constant in a dielectric medium. This topic has been a subject of active investigation, and in this respect we can 
quote e.g. a series of papers by Schwinger concerning a possible relation between 
dynamical Casimir effect (DCE) and sonoluminescence \cite{schwinger-sono}. In this paper, 
instead of starting, as Schwinger did, from the quantization 
of electrodynamics for a dielectric non dispersive medium, we refer to a less phenomenological situation in which dielectric properties 
are rooted into the interaction between electromagnetic field and a set of oscillators reproducing sources for dispersive properties 
of the electromagnetic field in matter, 
as in the well-known Hopfield model developed by Hopfield \cite{Hopfield,Fano,kittel,davydov}. We generalize the usual picture in the following sense: we work in a general framework for photon pair creation associated with a space-time dependent dielectric susceptibility, and in particular we focus our attention on perturbation theory.  
In \cite{Belgiorno-Hopfield} we have proposed a generalization of the so called Hopfield model, fully Lorentz covariant, and allowing the introduction of a
quite general class of spacetime dependent perturbations reproducing a multitude of physically interesting situations. We point out 
that we are implicitly assuming that absorption, which plays a fundamental role in Kramers-Kronig relation, is negligible. 
This assumption amounts to a first-step approximation, 
which is reasonable as far as field frequencies far from the resonances are considered, 
and as far as we focus on photon production induced by space-time dependent perturbations. See also the discussion in the following section.\\

 In this paper we perform a first order
perturbative analysis of the mentioned class, in order to investigate the induced photon pair production from vacuum. 
In particular, we will compute
the $S$ matrix element associated to the transition amplitude from the vacuum to a photon pair state. We will consider the case of a general dispersive
but non dissipative linear medium, with an arbitrary number of resonances, and will determine the number of photons emitted, as well as the number of
photon pairs produced by the presence of a time varying perturbation. 
Our theoretical picture appears to be applicable to 
several physically interesting situations where an intense laser pulse, shot into a nonlinear dielectric medium, generates 
a travelling dielectric perturbation thanks to the Kerr effect \cite{boyd}. We point out that, in our model, nonlinearity 
is phenomenologically taken into account simply through its effect, i.e. the presence of a refractive index perturbation 
travelling in the medium.  
A perturbative phenomenological approach 
is e.g. at the root of an interesting pair-creation phenomenon which displays a threshold for photon pair-creation depending on the velocity of the 
perturbation \cite{belgio-prl,DallaPiazza:2012aj}.  Our example refers just to this kind of travelling perturbation with 
constant velocity, which can be amenable of experimental set-up and verification, and represents an improvement of  
\cite{belgio-prl,DallaPiazza:2012aj}. Beyond the aforementioned phenomenology associated with the Kerr effect, we mention that also 
sonoluminescence could be taken into account in our framework (perturbation theory 
was applied in \cite{SPS}). A further interesting situation, where photon pair production 
is induced by a pulse with orbital angular momentum, will be described elsewhere \cite{Helix}.

We also underline that the present picture, at least on the side of dielectric 
properties of the medium, provides a coherent foundation and generalization of the results presented in \cite{SPS}. Indeed, 
a more fundamental setting for the theory is provided, and dispersive properties are automatically taken into account. Moreover, the 
possibility to obtain in an easier way higher order contributions is also given. 
 It is worth mentioning that in \cite{suttorp-jpa}, a very general picture and interesting picture is provided, 
where inhomogeneities with generic spatio-temporal dependence are allowed, 
and the susceptibility is a tensor field depending 
on space and time. Moreover, therein absorption is included by means of a bath of oscillators whose 
interactions with the electromagnetic field originate dissipative effects. Even if, in this respect, our model 
can be considered as a sub-case, holding for negligible absorption and for perturbative inhomogeneities, 
of this general approach, we point out that we develop a formalism leaving room for   
covariance and quantization in a covariant gauge, which are not treated therein.

\section{The Hopfield model}

Quantization in a dispersive medium can be approached in different ways. A possibility is to perform a quantization 
of the electromagnetic field by taking into account spacetime and frequency dependence of the dielectric constant and 
magnetic permeability. For purely dispersive effects, see e.g. \cite{milonni,ginzburg}. Alternatively, one can start from a less phenomenological 
picture, as in \cite{Hopfield,Fano,suttorp-jpa,suttorp,suttorp-wubs}. 
See also the recent monograph \cite{luks} for a survey on methods of 
quantization both in a phenomenological framework and in a microscopically grounded one. 
In particular, 
in \cite{Belgiorno-Hopfield} we have proposed a generalized relativistic covariant Hopfield model for the electromagnetic field in a dielectric dispersive medium
in a framework in which one allows a space-time dependent susceptibility, aimed to a phenomenological description of a space-time varying dielectric perturbation
induced by a local time dependent variation of the dielectric susceptibility. This is {\sl per s\`e} an interesting contribution to the microscopically-grounded works on the subject, because covariance and constrained quantization coexist and are 
coherently discussed. Covariance, as is known, and is confirmed since the original work by Minkowski \cite{minkowski} 
and e.g. by \cite{post,haus},
is not simply a speculative exercise in the picture at hand, but allows to 
get the correct behavior of physical quantities when changing from an inertial observer to another one. This e.g. is relevant in the discussion of the analogue of the Hawking effect (for the optical case, see e.g. 
\cite{petev-prl,rubino-njp,finazzi-hopfield,finazzi-wh,belgiorno-prd}), 
where passing to the reference frame 
which is comoving with the uniformly travelling perturbation is of basic relevance in order to understand several 
theoretical questions \footnote{We shall discuss the Hawking effect in the framework of the Hopfield model in 
a dedicated paper.}. 
Constrained quantization is as well an important topic for understanding the role of 
constraints on the quantization of the model at hand. See e.g. \cite{gitman-tyutin,henneaux-teitelboim}. 
Introducing absorbption as in \cite{suttorp,suttorp-wubs,suttorp-jpa} would make quite trickier both the constrained quantization 
scheme and the effective computations. We mean to come back on this topic in future works. 

In this paper, we follow a different strategy with respect to \cite{Belgiorno-Hopfield}, where no reduction
of the first-class constraints to second-class ones occurs. As a consequence, the Lorentz-Landau gauge condition we fix (see below) is
to be imposed by means of a Gupta-Bleuler condition on the physical states. Moreover, we use MKS unrationalized system.\\
In terms of the four-potential gauge field $A$ and a single polarization field $P$, it is described by the classical Hamiltonian density
\begin{align}
\ham &= \frac 12 (\Pi_A^i)^2 + \frac{1}{4} F_{ij} F^{ij} + A_0 (\partial_i \Pi_{A\; i}) \cr
&+\frac{1}{c} (v_0 P_i-v_i P_0)   \Pi_{A\; i} - \frac{v^k}{v_0}
(\pa_k P^\mu) \Pi_{P\; \mu} \cr
&-\frac{\chi}{2 (v^0)^2} \Pi_{P\; \mu} \Pi_P^\mu- \frac{\omega_0^2}{2\chi} P_\mu P^\mu \cr
&+\frac{1}{2c^4} (v_0 P_i-v_i P_0)^2,
\end{align}
$v^\mu$ is the four-velocity of the dielectric medium. 
The polarization field must satisfy the following condition
\beq
v^\mu P_\mu =0. \label{transverse}
\eeq
The space of complexified fields $(A,P)$ is endowed with the conserved scalar product
\begin{align}
&\pmb((A_\mu, P_\mu)\pmb |(\tilde A_\mu, \tilde P_\mu )\pmb )= \frac ic \int_{\Sigma_t} \left[F^{*0\nu} \tilde A_\nu
+\frac 1\chi v^\rho \partial_\rho P^{*\sigma} \tilde P_\sigma v^0+ \right. \cr
&\left. -\frac 1{c} (P^{*0} v^\rho -P^{*\rho} v^0) \tilde A_\rho
 -\tilde F^{0\nu} A^*_\nu -\frac 1\chi v^\rho \partial_\rho \tilde P^{\sigma} P^*_\sigma v^0 +\right. \cr
 &\left. + \frac
1{c} (\tilde P^{0} v^\rho -\tilde P^{\rho} v^0) A^*_\rho \right]d^3x.
\end{align}
This provide a natural structure for the procedure of quantization. Because of the presence of constraints this requires some carefulness,
and the result is that, beyond the standard CCR for the $A$ field, the correct CCR for the field $P^\mu$ and its conjugate momentum $\Pi^\mu$ is \cite{Belgiorno-Hopfield}:
\beq
[P^\mu,\Pi_P^\nu ] =
i \hbar \left(\eta^{\mu \nu} - \frac{1}{v_\rho v^\rho} v^\mu v^\nu\right) \delta^{(3)}
(\mathbf{x}-\mathbf{y}).
\eeq
Accordingly to the classical constraint condition $\partial_\mu A^\mu=0$, one imposes the following condition on the
physical states:
\beq
\partial_\mu A^{\mu\ (+)} |\Psi_{phys} \rangle = 0, \label{gupta-bleuler}
\eeq
where, with standard notation, we mean that the positive frequency (annihilator) part of the operator vanishes on the
physical states $ |\Psi_{phys} \rangle $. The further constraint  $v_\mu P^\mu=0$ is not related to a gauge invariance and
is automatically implemented \cite{Belgiorno-Hopfield}.\\
We will now consider the perturbative quantization of this model, by considering as unperturbed the model with constant susceptibility
coefficient $\chi(\mathbf{x},t)=\chi_0$. The perturbation will then be parameterized by the function
$\delta\chi(\mathbf{x},t)=\chi(\mathbf{x},t)-\chi_0$.

\section{Perturbative on-shell quantization of the Hopfield model}
\begin{figure}[htbp]
\centering
\includegraphics[width=9cm]{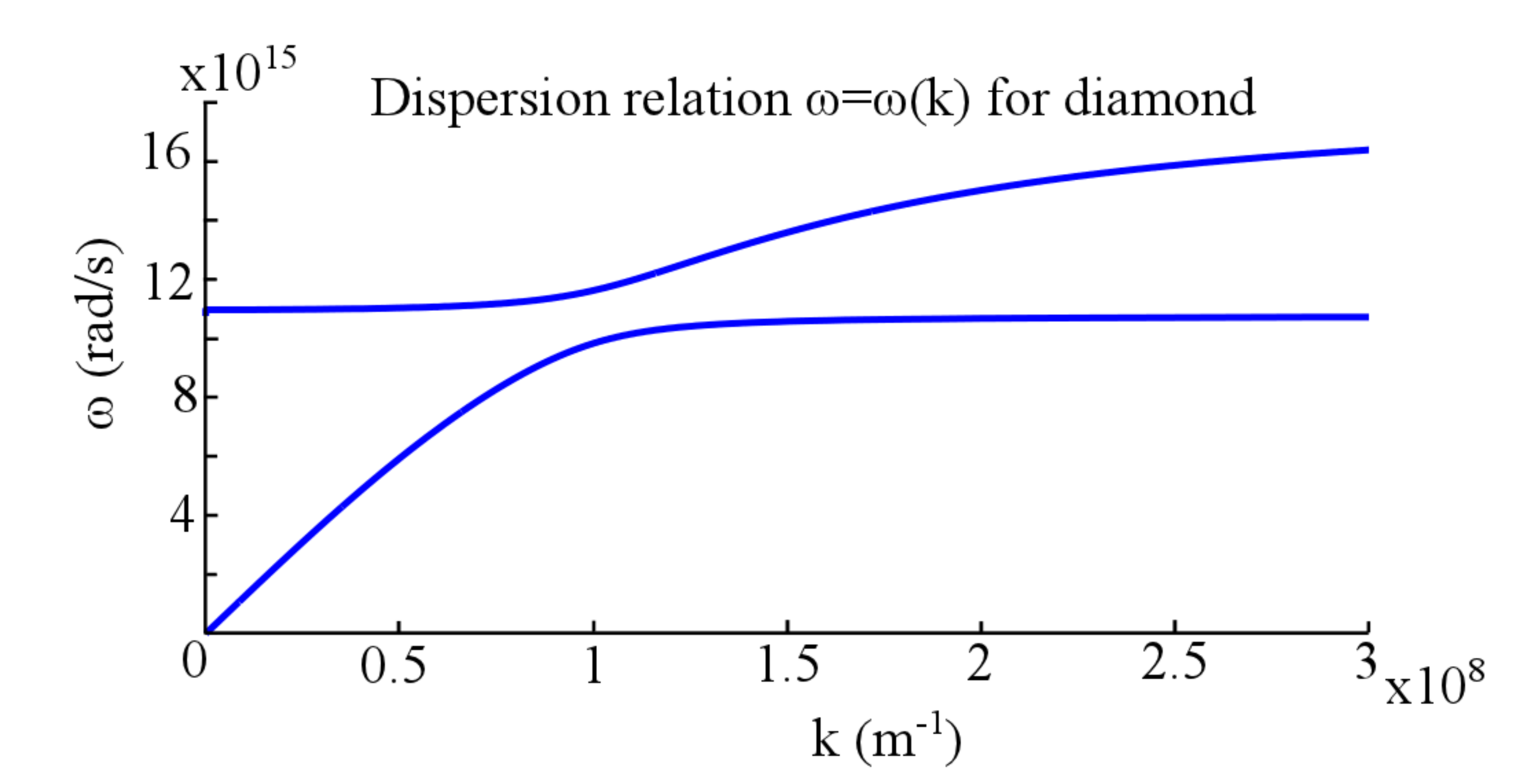}
\caption{The two lowest dispersion branches for diamond. The third one appears for $\omega > 4\cdot 10^{16}\, \text{rad/s}$.}
\label{fig:disp}
\end{figure}

We consider unperturbed the case of constant $\chi_0$ \cite{Hopfield,Fano}, which is exactly tractable.
The exact solutions of the classical equations of motion in the lab frame, and in the Lorentz-Landau gauge, are generated by plane waves having only
spatial components
\begin{eqnarray}
&& \mathbf {A}(\mathbf {x},t)=\mathbf {A}_0 e^{-i\omega t +i \mathbf {k}\cdot \mathbf {x}},\\
&& \mathbf {P}(\mathbf {x},t)=-i \frac {\chi_0 \omega}{\omega_0^2-\omega^2}\frac 1c \mathbf {A}_0 e^{-i\omega t +i \mathbf {k}\cdot \mathbf {x}},\label{condition}
\end{eqnarray}
where $\omega$ is related to $\mathbf k$ by the dispersion relation
\begin{eqnarray}
c^2\mathbf{k}^2 = \omega^2 \left[ 1+\frac {\chi_0}{\omega_0^2-\omega^2} \right]=\omega^2 n^2_p(\omega), \label{dispersion}
\end{eqnarray}
where $n_p(\omega)$ is the phase velocity refractive index.
The last relation selects two branches separated by the resonance frequency $\pm \omega_0$, the first and the second dispersion branches. Figure \ref{fig:disp} concerns a more general case (cf. section \ref{resonances}) of two resonances, as for diamond.
We call
$\pm\omega_-$ the modes in the first dispersion branch, and $\pm \omega_+$ the solutions in the other branch. Thus, we have
\begin{align}
\omega_\pm^2 &=\frac {\omega_0^2+\chi_0+c^2 \mathbf {k}^2}2 + \cr
&\pm \frac 12 \sqrt {[(|\mathbf {k}|c+\omega_0)^2+\chi_0][(|\mathbf
{k}|c-\omega_0)^2+\chi_0]}.
\label{d-branches}
\end{align}
It is easy to show that indeed $\omega_-<\omega_0$ and $\omega_+>\omega_0$. The general solution for the $\mathbf A$ field
can be written in the form
\begin{align}
A^j(\mathbf {x},t)&=\int d\omega \int \frac {d^3\mathbf k}{(2\pi)^3} [f^j(\omega, \mathbf{k}) e^{-i\omega t +i \mathbf {k}\cdot \mathbf
{x}} \cdot \cr
&\cdot \delta(\mathbf{k}^2 - \frac {\omega^2}{c^2}\left[ 1+\frac {\chi_0}{\omega_0^2-\omega^2} \right]) +c.c.].
\end{align}
We can now integrate explicitly in $\omega$ by employing the properties of the $\delta$ function. This gives
\begin{align}
{A^j}(\mathbf {x},t)&=\int \frac {d^3\mathbf k}{\Phi^-_{\mathbf k}} [a^j_{\mathbf k} e^{-i\omega_- t +i \mathbf {k}\cdot \mathbf
{x}} +a^{j*}_{\mathbf k} e^{i\omega_- t -i \mathbf {k}\cdot \mathbf {x}}] +\cr
&+\int \frac {d^3\mathbf k}{\Phi^+_{\mathbf k}} [\tilde
a^j_{\mathbf k} e^{-i\omega_+ t +i \mathbf {k}\cdot \mathbf {x}} +\tilde a^{j*}_{\mathbf k} e^{i\omega_+ t -i \mathbf {k}\cdot \mathbf {x}}],
\end{align}
where we have introduced the measure factor (coming from the $\delta$ distribution)
\begin{align}
\Phi^\pm_{\mathbf k} &= \frac 2{c^2} \omega_\pm (2\pi)^3 \left[ 1+\frac {\chi_0 \omega_0^2}{(\omega_0^2-\omega_\pm^2)^2} \right] \cr
&=
\frac 2{c^2} \omega_\pm (2\pi)^3 n_g(\omega_\pm) n_p(\omega_\pm),
\end{align}
where $n_g(\omega)$ is the group velocity refractive index.
Note that the field results to be naturally the sum of $\omega_-$ modes with amplitude $a_{\mathbf k}$ and $\omega_+$ modes with amplitude
$\tilde a_{\mathbf k}$.\\
In the same way we can compute the polarization field and the associated momenta:
\begin{align}
& {P^j}(\mathbf {x},t)=-\frac ic \int \frac {d^3\mathbf k}{\Phi^-_{\mathbf k}} \frac {\chi_0 \omega_-}{\omega_0^2-\omega_-^2} [a^j_{\mathbf k} e^{-i\omega_- t +i
\mathbf {k}\cdot \mathbf {x}} + \cr
&\phantom{{P^i}(\mathbf {x},t)=} -a^{j*}_{\mathbf k} e^{i\omega_- t -i \mathbf {k}\cdot \mathbf {x}}] + \cr
&\phantom{{P^i}(\mathbf {x},t)=}
-\frac ic \int \frac {d^3\mathbf k}{\Phi^+_{\mathbf k}} \frac {\chi_0 \omega_+}{\omega_0^2-\omega_+^2}
[\tilde a^j_{\mathbf k} e^{-i\omega_+ t +i \mathbf {k}\cdot \mathbf {x}} + \cr
&\phantom{{P^i}(\mathbf {x},t)=} -\tilde a^{j*}_{\mathbf k} e^{i\omega_+ t -i \mathbf {k}\cdot \mathbf {x}}],\\
& {\Pi^j_A}(\mathbf {x},t)=\frac i{c^2} \int \frac {d^3\mathbf k}{\Phi^-_{\mathbf k}} \omega_- [a^j_{\mathbf k} e^{-i\omega_- t +i
\mathbf {k}\cdot \mathbf {x}} -a^{j*}_{\mathbf k} e^{i\omega_- t -i \mathbf {k}\cdot \mathbf {x}}]+ \cr 
&\phantom{{\Pi^i_A}(\mathbf
{x},t)=} +\frac i{c^2} \int \frac {d^3\mathbf k}{\Phi^+_{\mathbf k}} \omega_+ [\tilde a^j_{\mathbf k}
e^{-i\omega_+ t +i \mathbf {k}\cdot \mathbf {x}} + \cr
&\phantom{{\Pi^i_A}(\mathbf
{x},t)=} -\tilde a^{j*}_{\mathbf k} e^{i\omega_+ t -i \mathbf {k}\cdot \mathbf {x}}] -\frac 1c {P^i}(\mathbf {x},t), \\
& {\Pi^j_{P}}(\mathbf {x},t)=\frac 1c \int \frac {d^3\mathbf k}{\Phi^-_{\mathbf k}} \frac {\omega_-^2}{\omega_0^2-\omega_-^2}
[a^j_{\mathbf k} e^{-i\omega_- t +i \mathbf {k}\cdot \mathbf {x}} + \cr
&\phantom{{\Pi^i_{P}}(\mathbf {x},t)=} +a^{j*}_{\mathbf k} e^{i\omega_- t -i \mathbf {k}\cdot \mathbf
{x}}] + \cr 
&\phantom{{\Pi^i_P}(\mathbf {x},t)=}+\frac 1c \int \frac {d^3\mathbf k}{\Phi^+_{\mathbf k}} \frac
{\omega_+^2}{\omega_0^2-\omega_+^2} [\tilde a^j_{\mathbf k} e^{-i\omega_+ t +i \mathbf {k}\cdot \mathbf {x}} +\cr
&\phantom{{\Pi^i_P}(\mathbf {x},t)=} +\tilde
a^{j*}_{\mathbf k} e^{i\omega_+ t -i \mathbf {k}\cdot \mathbf {x}}].
\end{align}
In the Hamiltonian formulation, $\mathbf {A}, \mathbf {P}, \mathbf {\Pi}_A, \mathbf {\Pi}_P$ are the dynamical variables subject to a
canonical symplectic structure at fixed time, with non vanishing Poisson brackets
\begin{eqnarray}
&& \{ {A}^i(\mathbf {x}, t), {\Pi}^j_A (\mathbf {x}', t) \}=-\delta^{ij}\delta^3(\mathbf {x}-\mathbf {x}'), \label{our-canon-1}\\
&& \{ {P}^i(\mathbf {x}, t), {\Pi}^j_P (\mathbf {x}', t) \}=-\delta^{ij}\delta^3(\mathbf {x}-\mathbf {x}'),
\label{our-canon-2}
\end{eqnarray}
so that the Hamilton equations
\begin{eqnarray}
&& \partial_t \mathbf {A}=-\{H, \mathbf {A} \}, \qquad\qquad\ \partial_t \mathbf {\Pi}_A=-\{H, \mathbf {\Pi}_A \}, \\
&& \partial_t \mathbf {P}=-\{H, \mathbf {P} \}, \qquad\qquad\
\partial_t \mathbf {\Pi}_P=-\{H, \mathbf {\Pi}_P \},
\end{eqnarray}
are equivalent to the original Lagrange equations. One can proceed with quantization by promoting the dynamical variables to operators
and the Poisson brackets to commutators defined by the correspondence principle. Equivalently, we can use $a^i_{\mathbf k}, \tilde a^i_{\mathbf k}$
and their conjugates as new dynamical variables. We will use the same symbols for the corresponding operators.
Notice that, if we indicate with
\begin{eqnarray}
\mathcal {U}_\pm =\left( \mathbf {\xi} e^{-i\omega_\pm t+i\mathbf{k} \cdot \mathbf {x}}, -i \frac {\chi_0
\omega_\pm}{c(\omega_0^2-\omega_\pm^2)} \mathbf {\xi} e^{-i\omega_\pm t+i\mathbf{k} \cdot \mathbf {x}}\right)
\end{eqnarray}
the standard plane wave of amplitude $\mathbf{\xi}$, we find that
\begin{eqnarray}
&& \mathbf {\xi}^* \cdot \mathbf {a}_{\mathbf k}= \pmb ( \mathcal {U}_- , ( \mathbf {A}, \mathbf {P} ) \pmb ), \\
&& \mathbf {\xi} \cdot \mathbf {a}^\dagger_{\mathbf k}= \pmb ( \mathcal {U}^*_- , ( \mathbf {A}, \mathbf {P} ) \pmb ), \\
&& \mathbf {\xi}^* \cdot \mathbf {\tilde a}_{\mathbf k}= \pmb ( \mathcal {U}_+ , ( \mathbf {A}, \mathbf {P} ) \pmb ), \\
&& \mathbf {\xi} \cdot \mathbf {\tilde a}^\dagger_{\mathbf k}= \pmb ( \mathcal {U}^*_+ , ( \mathbf {A}, \mathbf {P} ) \pmb ).
\end{eqnarray}
and that the oscillators satisfy the canonical brackets
\begin{eqnarray}
&& [ a^i_{\mathbf k} , a^{j\dagger}_{\mathbf {k}'}]= (\delta^{ij}-\frac {k^ik^j}{\mathbf{k}^2}) \Phi^-_{\mathbf k} \delta^3 (\mathbf {k}-\mathbf {k}'),\\
&& [\tilde a^i_{\mathbf k} , \tilde a^{j\dagger}_{\mathbf {k}'}]= (\delta^{ij}-\frac {k^ik^j}{\mathbf{k}^2}) \Phi^+_{\mathbf k} \delta^3 (\mathbf {k}-\mathbf {k}'), \\
&& [a^i_{\mathbf k} , \tilde a^{j}_{\mathbf {k}'}]= 0, \qquad\qquad\ [a^i_{\mathbf k} , \tilde a^{j\dagger}_{\mathbf {k}'}]= 0, \\
&& [a^{i\dagger}_{\mathbf k} , \tilde a^{j}_{\mathbf {k}'}]= 0, \qquad\qquad\ [\tilde a^{i\dagger}_{\mathbf k} , \tilde a^{j\dagger}_{\mathbf {k}'}]=0.
\end{eqnarray}
However, recalling that the oscillator fields are constrained by the transversality condition, it is convenient
to consider unconstrained oscillating field operators $ a_{\mu\mathbf k}$, $\tilde a_{\mu\mathbf k}$, $\mu=1,2$ and express the fields in terms
of the constrained operators $\sum_\mu \mathbf {e}^*_{-\mu\mathbf k} {a}_{\mu\mathbf k}$, $\sum_\mu \mathbf {e}^*_{+\mu\mathbf k} \tilde a_{\mu\mathbf k}$,
where $\mathbf {e}_{\pm\mu\mathbf k}$, $\mu=1,2$ form two bases (one for each sign) of the polarization vectors orthogonal to $\mathbf{k}$, and
satisfying the relations
\begin{eqnarray}
\sum_\mu e^i_{\pm\mu\mathbf k} e^j_{\pm\mu\mathbf k}=\delta^{ij}-\frac {k^ik^j}{\mathbf{k}^2},
\end{eqnarray}
whereas the unconstrained operators satisfy
\begin{eqnarray}
&& [ a_{\mu\mathbf k} , a^{\dagger}_{\nu\mathbf {k}'}]= \delta_{\mu\nu} \Phi^-_{\mathbf k} \delta^3 (\mathbf {k}-\mathbf {k}'),\\
&& [\tilde a_{\mu\mathbf k} , \tilde a^{\dagger}_{\nu\mathbf {k}'}]= \delta_{\mu\nu} \Phi^+_{\mathbf k} \delta^3 (\mathbf {k}-\mathbf {k}'), \\
&& [a_{\mu\mathbf k} , \tilde a_{\nu\mathbf {k}'}]= 0, \qquad\qquad\ [a_{\mu\mathbf k} , \tilde a^{\dagger}_{\nu\mathbf {k}'}]= 0, \\
&& [a^{\dagger}_{\mu\mathbf k} , \tilde a_{\nu\mathbf {k}'}]= 0, \qquad\qquad\ [\tilde a^{\dagger}_{\mu\mathbf k} , \tilde a^{\dagger}_{\nu\mathbf {k}'}]=0.
\end{eqnarray}
The unperturbed Hamiltonian operator is defined via the normal ordered operator
\begin{align}
H_0=:\int d^3 \mathbf {x} &\left[\frac {c^2}2 \mathbf {\Pi}_A^2 -\frac 1{2} \mathbf {A} \cdot \Delta \mathbf {A} +c \mathbf {P}
\cdot \mathbf {\Pi}_A + \right. \cr
&\left.+\frac {\chi_0}2 \mathbf {\Pi}_P^2 +\frac 12\left( \frac {\omega_0^2}{\chi_0} -1 \right) \mathbf P^2 \right]:,
\end{align}
and expressed in terms of the oscillator operators takes the form
\begin{align}
H_0&= \sum_{\mu=1}^2 \int \frac {d^3 \mathbf{k}}{\Phi^-_{\mathbf k}} a^\dagger_{\mu\mathbf k}
a_{\mu\mathbf k} \ \hbar \omega_- + \cr
&+ \sum_{\mu=1}^2 \int \frac {d^3 \mathbf{k}}{\Phi^+_{\mathbf k}} \tilde a^\dagger_{\mu\mathbf k}
\tilde a_{\mu\mathbf k} \ \hbar \omega_+. \label{hamiltonian}
\end{align}
This allows to interpret
\begin{eqnarray}
\frac {d^3 \mathbf{k}}{\Phi^-_{\mathbf k}} a^\dagger_{\mu\mathbf k} a_{\mu\mathbf k}
\end{eqnarray}
as the number operator for the polaritons in the first branch, with energy $\hbar \omega_-$, wave vector in $\mathbf{k}-\mathbf {k}+
d^3 \mathbf {k}$, and polarization $\mathbf {e}_{-\mu\mathbf k}$, and similar for the second branch.\\
It may be noted that the Lorentz-Landau gauge we imposed at the beginning of our calculations, due to the equations 
of motion at the unperturbed level, still lead us to $A_0^{(0)}=0$ (the upper index 
indicates the order in the perturbative expansion), and then, at least at the unperturbed level, we find 
again standard transversality occurring in the Coulomb gauge. As our inhomogeneous perturbation plays the role 
of source for the divergence of the polarization field, we expect that such a transvesality is broken at higher 
order, leaving us with the necessity of a Gupta-Bleuler formalism. 

\subsection{Generalization to an arbitrary number of resonance frequencies} \label{resonances}
Consider the case of $N>1$ material harmonic oscillators coupled with the electromagnetic field. These can be described
by the Hamiltonian:
\begin{align}
\ham_{N} &= \frac 12 (\Pi_A^i)^2 + \frac{1}{4} F_{ij} F^{ij} + A_0 (\partial_i \Pi_{A\; i}) \cr
&+\sum_{k=1}^N \left[\frac{1}{c} (v_0 P_{(k) i}-v_i P_{(k) 0})   \Pi_{A\; i} - \frac{v^i}{v_0}
(\pa_i P^\mu_{(k)}) \Pi_{P_{(k)} \mu} \right. \cr
&\left. -\frac{\chi}{2 (v^0)^2} \Pi_{P_{(k)} \mu} \Pi_{P_{(k)}}^\mu- \frac{\omega_{(k) 0}^2}{2\chi} P_{(k) \mu} P_{(k)}^\mu \right.\cr
&\left. +\frac{1}{2c^4} (v_0 P_{(k) i}-v_i P_{(k) 0})^2\right].
\end{align}
The quantum fields $P^\mu_{(k)}$ satisfy
\beqnl
\left[P^\mu_{(k)},\Pi_{P^\nu_{(l)}}\right] :&=& i\hbar \delta_{(k) (l)}
\left(\eta^{\mu \nu} - \frac{v^\mu v^\nu }{v_\rho v^\rho} \right) \delta^3 (\mathbf{x}-\mathbf{y}),\\
\left[P^\mu_{(k)}, P^\nu_{(l)}\right] :&=& 0,\\
\left[\Pi_{P^\mu_{(k)}},\Pi_{P^\nu_{(l)}}\right]:&=& 0.
\eeqnl
In the case when $\chi_{(k)}=\chi_{0(k)}$ are constant, the classical equation of motion can be solved exactly by using the Fourier transform method.
The solutions result to be governed by the dispersion relation
\begin{eqnarray}
c^2\mathbf{k}^2 = \omega^2 \left[ 1+\sum_{l=1}^N\frac {\chi_{0(l)}}{\omega_{0(l)}^2-\omega^2} \right]=\omega^2 n^2_p(\omega). \label{dispersion-N}
\end{eqnarray}
It is easy to see that for any value of $\mathbf{k}^2$ this equation admits $N+1$ positive solutions $\omega^2_{\alpha\mathbf{k}}$,
$\alpha=0,1,\ldots, N$ corresponding to $N+1$ dispersion branches, satisfying $\omega^2_{\alpha\mathbf{k}}< \omega_{0(\alpha+1)}^2 < \omega^2_{\alpha+1\mathbf{k}}$,
$\alpha=0,\ldots,N-1$. \\
Again, we can introduce polarization vectors $e^i_{\alpha\mu\mathbf k}$, $\mu=1,2$, $\alpha=0,1,\ldots,N$ satisfying
\begin{eqnarray}
\sum_\mu e^i_{\alpha\mu\mathbf k} e^j_{\alpha\mu\mathbf k}=\delta^{ij}-\frac {k^ik^j}{\mathbf{k}^2},
\end{eqnarray}
so that the fields take the form
\begin{align}
&\mathbf{A}(\mathbf{x},t)=\sum_{\mu=1}^2 \sum_{\alpha=0}^N \int \frac {d^3\mathbf{k}}{\Phi^{\alpha}_{\mathbf{k}}} \left[ \mathbf {e}^*_{\alpha\mu\mathbf k}
a_{\mu\alpha\mathbf{k}} e^{-i\omega_{\alpha\mathbf{k}}t+i\mathbf{k}\cdot \mathbf{x}} +\right. \cr
&\left. +\mathbf {e}_{\alpha\mu\mathbf k} a_{\mu\alpha\mathbf{k}}^{\dagger}
e^{i\omega_{\alpha\mathbf{k}}t-i\mathbf{k}\cdot \mathbf{x}}\right], \\
& \mathbf{P}_{(l)}(\mathbf{x},t)=-\frac {i}c \sum_{\mu=1}^2 \sum_{\alpha=0}^N \int \frac {d^3\mathbf{k}}{\Phi^{\alpha}_{\mathbf{k}}}
\frac {\chi_{0(l)} \omega_{\alpha\mathbf{k}}}{\omega_{0(l)}^2-\omega^2_{\alpha\mathbf{k}}} \cdot \cr
& \cdot \left[ \mathbf {e}^*_{\alpha\mu\mathbf k} a_{\mu\alpha\mathbf{k}}
e^{-i\omega_{\alpha\mathbf{k}}t+i\mathbf{k}\cdot \mathbf{x}} -\mathbf {e}_{\alpha\mu\mathbf k} a_{\mu\alpha\mathbf{k}}^{\dagger}
e^{i\omega_{\alpha\mathbf{k}}t-i\mathbf{k}\cdot \mathbf{x}}\right],
\end{align}
where we have introduced the invariant measure factors
\begin{align}
\Phi^{\alpha}_{\mathbf{k}}&=\frac 2{c^2} \omega_{\alpha\mathbf{k}} (2\pi)^3 \left[ 1+\sum_{l=1}^N \frac {\chi_{0(l)}
\omega^2_{0(l)}}{(\omega^2_{0(l)}-\omega^2_{\alpha\mathbf{k}})^2} \right]\cr
&=\frac 2{c^2} \omega_{\alpha\mathbf{k}}
(2\pi)^3 n_g(\omega_{\alpha\mathbf{k}}) n_p(\omega_{\alpha\mathbf{k}}).
\end{align}
The oscillator field operators satisfy the canonical commutators
\begin{align}
&[ a_{\mu\alpha\mathbf k} , a^{\dagger}_{\nu\beta\mathbf {k}'}]= \delta_{\mu\nu} \delta_{\alpha\beta} \Phi^{\alpha}_{\mathbf k} \delta^3 (\mathbf {k}-\mathbf {k}'), \cr 
&[ a_{\mu\alpha\mathbf k} , a_{\nu\beta\mathbf {k}'}]=[ a^{\dagger}_{\mu\alpha\mathbf k} , a^{\dagger}_{\nu\beta\mathbf {k}'}]=0.
\end{align}
The unperturbed Hamiltonian is
\begin{eqnarray}
H_0= \sum_{\mu=1}^2 \sum_{\alpha=0}^N\int \frac {d^3 \mathbf{k}}{\Phi^{\alpha}_{\mathbf k}} a^\dagger_{\mu\alpha\mathbf k} a_{\mu\alpha\mathbf k}\
\hbar \omega_{\alpha\mathbf{k}}.
\end{eqnarray}
This allows to interpret $a^\dagger_{\mu\alpha\mathbf k} a_{\mu\alpha\mathbf k}$ as a number density so that
\begin{eqnarray}
\frac {d^3 \mathbf{k}}{\Phi^{\alpha}_{\mathbf k}} a^\dagger_{\mu\alpha\mathbf k} a_{\mu\alpha\mathbf k}
\end{eqnarray}
is the number operator for polaritons with energy $\hbar \omega_{\alpha\mathbf{k}}$, wave vector in $\mathbf{k}-\mathbf{k}+d\mathbf{k}$,
and polarization $\mathbf {e}_{\alpha\mu\mathbf k}$.

\subsection{Photon emission induced by a perturbation}
The simplest perturbation of the system can be obtained by changing
\begin{eqnarray}
\chi_{0(l)}\rightarrow \chi_{(l)}(\mathbf{x},t)=\chi_{0(l)}+\delta \chi_{(l)}(\mathbf{x},t).
\end{eqnarray}
Then, the Hamiltonian is perturbed by a term
\begin{align}
\delta H=\sum_{l=1}^N\int &\left[\frac 12 \delta \chi_{(l)} \mathbf {\Pi}^2_{\mathbf {P}(l)} + \right. \cr
&\left. +  \frac 12 \omega^2_{0(l)} \left( \frac 1{\chi_{(l)}}
-\frac 1{\chi_{0(l)}} \right)\mathbf {P}^2_{(l)} \right]
d^3 \mathbf{x},
\end{align}
and using the expressions of the field $\mathbf{P}$ and its conjugate momentum $\mathbf{\Pi}_{\mathbf P}$ in terms of the oscillating modes
\begin{align}
&\mathbf{P}_{(l)}(\mathbf{x},t)=-\frac {i}c \sum_{\mu=1}^2 \sum_{\alpha=0}^N \int \frac {d^3\mathbf{k}}{\Phi^{\alpha}_{\mathbf{k}}}
\frac {\chi_{0(l)} \omega_{\alpha\mathbf{k}}}{\omega_{0(l)}^2-\omega^2_{\alpha\mathbf{k}}} \cdot \cr
& \cdot \left[ \mathbf {e}^*_{\alpha\mu\mathbf{k}} a_{\mu\alpha\mathbf{k}}
e^{-i\omega_{\alpha\mathbf{k}}t+i\mathbf{k}\cdot \mathbf{x}} -\mathbf {e}_{\alpha\mu\mathbf{k}} a_{\mu\alpha\mathbf{k}}^{\dagger}
e^{i\omega_{\alpha\mathbf{k}}t-i\mathbf{k}\cdot \mathbf{x}}\right],\\
& \mathbf {\Pi}_{\mathbf {P}(l)}(\mathbf {x},t)=\frac 1c \sum_{\alpha=0}^N \int \frac {d^3\mathbf k}{\Phi^{\alpha}_{\mathbf k}}
\frac {\omega_{\alpha\mathbf{k}}^2}{\omega_{0(l)}^2-\omega_{\alpha\mathbf{k}}^2} \cdot \cr
& \cdot [\mathbf {e}^*_{\alpha\mu\mathbf{k}} a_{\mu\alpha\mathbf k} e^{-i\omega_{\alpha\mathbf{k}} t +i \mathbf {k}\cdot \mathbf {x}}
+\mathbf {e}_{\alpha\mu\mathbf{k}} a^{\dagger}_{\mu\alpha\mathbf k}
e^{i\omega_{\alpha\mathbf{k}} t -i \mathbf {k}\cdot \mathbf {x}}],
\end{align}
we obtain:
\begin{align}
\delta H({\mathbf x}&,t)=
\frac{\hbar}{2c^2}\sum_{l=1}^N \sum_{\mu=1}^2 \left\{
\left[\sum_{\alpha=0}^N \int\frac {d^3\mathbf k}{\Phi^\alpha_{\mathbf k}} \frac {\omega_{\alpha {\mathbf k}}^2}{\omega_{0(l)}^2-\omega_{\alpha {\mathbf k}}^2} \cdot \right. \right. \cr
&\left.\left. \cdot (a_{\mu\alpha \mathbf k} e^{-i\omega_{\alpha {\mathbf k}} t +i \mathbf {k}\cdot \mathbf {x}} + a^{\dagger}_{\mu\alpha {\mathbf k}} e^{i\omega_{\alpha {\mathbf k}} t
-i \mathbf {k}\cdot \mathbf {x}})\right]\cdot \right. \cr
& \cdot \left[\sum_{\beta=0}^N \int\frac {d^3\mathbf k'}{\Phi^\beta_{\mathbf k'}} \frac {\omega_{\beta {\mathbf k'}}^2}{\omega_{0(l)}^2-\omega_{\beta {\mathbf k'}}^2}
(a_{\mu\beta \mathbf k'} e^{-i\omega_{\beta {\mathbf k'}} t +i \mathbf {k'}\cdot \mathbf {x}}  + \right. \cr
&\left. +a^{\dagger}_{\mu\beta {\mathbf k'}} e^{i\omega_{\beta {\mathbf k'}} t
-i \mathbf {k'}\cdot \mathbf {x}})\right] \delta\chi(\mathbf{x},t) \cr
& \dot\, \left[\sum_{\beta=0}^N \int\frac {d^3\mathbf k'}{\Phi^\beta_{\mathbf k'}} \frac {\omega_{\beta {\mathbf k'}}^2}{\omega_{0(l)}^2-\omega_{\beta {\mathbf k'}}^2}
(a_{\mu\beta \mathbf k'} e^{-i\omega_{\beta {\mathbf k'}} t +i \mathbf {k'}\cdot \mathbf {x}}  + \right. \cr
&\left. +a^{\dagger}_{\mu\beta {\mathbf k'}} e^{i\omega_{\beta {\mathbf k'}} t
-i \mathbf {k'}\cdot \mathbf {x}})\right] \delta\chi(\mathbf{x},t) \cr
&
+\omega_{0(l)}^2 \left( \frac{1}{\chi_{0(l)}+\delta\chi(\mathbf{x},t)} -\frac{1}{\chi_{0(l)}} \right)\cdot\cr
&\cdot \left[\sum_{\alpha=0}^N\int
\frac {d^3\mathbf k}{\Phi^\alpha_{\mathbf k}} \frac {\chi_{0(l)} \omega_{\alpha{\mathbf k}}}{\omega_{0(l)}^2-\omega_{\alpha{\mathbf k}}^2}
(a_{\mu\alpha{\mathbf k}} e^{-i\omega_{\alpha{\mathbf k}} t +i \mathbf {k}\cdot \mathbf {x}} +\right. \cr
&\left. -a^{\dagger}_{\mu\alpha{\mathbf k}} e^{i\omega_{\alpha{\mathbf k}} t
-i \mathbf {k}\cdot \mathbf {x}})\right] \cdot  \cr
&\left.
\cdot \left[\sum_{\beta=0}^N\int \frac {d^3\mathbf k'}{\Phi^\beta_{\mathbf k'}} \frac {\chi_{0(l)} \omega_{\beta{\mathbf k'}}}{\omega_{0(l)}^2
-\omega_{\beta{\mathbf k'}}^2} \cdot \right. \right.\cr
&\left. \left. \cdot (a_{\mu\beta{\mathbf k'}} e^{-i\omega_{\beta{\mathbf k'}} t +i \mathbf {k'}\cdot \mathbf {x}} -a^{\dagger}_{\mu\beta{\mathbf k'}}
e^{i\omega_{\beta{\mathbf k'}} t -i \mathbf {k'}\cdot \mathbf {x}})\right] \right\}.
\end{align}
Let us compute the probability amplitude for creating a pair of polaritons, the first one in the branch $\alpha_1$, with wave vector $\mathbf {k}_1$ and polarization
$\pmb \zeta_{\mathbf {k}_1}=\sum_{\nu=1}^2 \zeta_\nu \mathbf {e}_{\alpha_1 \nu \mathbf{k}_1}$, and the second one in the branch $\alpha_2$, with wave vector
$\mathbf {k}_2$ and polarization $\pmb \xi_{\mathbf {k}_2}=\sum_{\rho=1}^2 \xi_\rho \mathbf {e}_{\alpha_2 \rho \mathbf{k}_2}$. This corresponds to the state
\begin{eqnarray}
| \alpha_1 \pmb {\zeta} \mathbf{k}_1; \alpha_2 \pmb {\xi} \mathbf{k}_2 \rangle= \sum_{\nu=1}^2 \sum_{\rho=1}^2 \zeta_{\nu} \xi_\rho
a^\dagger_{\nu\alpha_1\mathbf {k}_1} a^\dagger_{\rho\alpha_2\mathbf {k}_2} |0\rangle.
\end{eqnarray}
This is given by
\begin{eqnarray}
\mathcal {A}_{\{ \alpha_1 \pmb {\zeta} \mathbf{k}_1; \alpha_2 \pmb {\xi} \mathbf{k}_2 \}}=
\langle \alpha_1 \pmb {\zeta} \mathbf{k}_1; \alpha_2 \pmb {\xi} \mathbf{k}_2 | S |0 \rangle,
\end{eqnarray}
where at first order the $S$-matrix is given by
\begin{eqnarray}
S\simeq\mathbb {I} -\frac i{\hbar} \int d^3 \mathbf {x} dt \delta H(\mathbf {x}, t).
\end{eqnarray}
At this order we can approximate
\begin{eqnarray}
\frac{1}{\chi_{0(l)}+\delta\chi(\mathbf{x},t)} -\frac{1}{\chi_{0(l)}}\simeq -\frac{\delta\chi(\mathbf{x},t)}{\chi_{0(l)}^2},
\end{eqnarray}
so that we get
\begin{align}
&\mathcal {A}_{\{ \alpha_1 \pmb {\zeta} \mathbf{k}_1; \alpha_2 \pmb {\xi} \mathbf{k}_2 \}}
=-\frac i{2c^2} \sum_{\mu=1}^2 \sum_{\nu=1}^2 \sum_{\rho=1}^2 \sum_{l=1}^N \sum_{\alpha=0}^N \sum_{\beta=0}^N \cr
&\int d^3 \mathbf {x} dt \int \frac {d^3 \mathbf k}{\Phi^\alpha_{\mathbf{k}}} \int \frac {d^3 \mathbf k'}{\Phi^\beta_{\mathbf{k'}}}
\left\{ \frac {\omega_{\alpha\mathbf k}\omega_{\beta\mathbf {k'}}(\omega_{\alpha\mathbf k}\omega_{\beta\mathbf {k'}}
+\omega^2_{0(l)})}{(\omega^2_{0(l)}-\omega_{\alpha\mathbf k}^2)(\omega^2_{0(l)}-\omega_{\beta\mathbf {k'}}^2)} \cdot \right. \cr
&\left. \cdot e^{it(\omega_{\alpha\mathbf {k}}+\omega_{\beta\mathbf {k'}})-i\mathbf {x}\cdot (\mathbf {k}+\mathbf {k'})}
\delta \chi_{(l)} (\mathbf {x}, t)\cdot \right. \cr
&\left. \cdot \langle 0 | \zeta^*_{\nu} a_{\nu\alpha_1\mathbf {k}_1} \xi^*_\rho a_{\rho\alpha_2\mathbf {k}_2}
a^\dagger_{\mu\alpha\mathbf {k}} a^\dagger_{\mu\beta\mathbf {k'}} | 0 \rangle \right\} \cr
&= \frac 1{c^2} \pmb \zeta \cdot \pmb \xi\ \sum_{l=0}^N
\frac {\omega_{\alpha_1\mathbf k_1}\omega_{\alpha_2\mathbf {k}_2}(\omega_{\alpha_1\mathbf k_1}\omega_{\alpha_2\mathbf {k}_2}
+\omega^2_{0(l)})}{(\omega^2_{0(l)}-\omega_{\alpha_1\mathbf k_1}^2)(\omega^2_{0(l)}-\omega_{\alpha_2\mathbf {k}_2}^2)} \cdot\cr
&\cdot \widehat {\delta\chi}_{(l)} (\omega_{\alpha_1\mathbf {k}_1}+\omega_{\alpha_2\mathbf {k}_2}, \mathbf {k}_1+\mathbf {k}_2),
\end{align}
where
\begin{eqnarray}
\widehat {\delta\chi}_{(l)} (\omega_{\alpha\mathbf {k}}, \mathbf {k})=
\int d^3 \mathbf {x} dt\ e^{i\omega_{\alpha\mathbf {k}}t -i\mathbf {k} \cdot \mathbf x} \delta\chi_{(l)} (\mathbf x, t).
\end{eqnarray}
From this we can compute the number of polaritons generated with wave vector $\vec k$ in the solid angle $d\omega_{\mathbf k}$,
in the branch $\alpha$, with polarization $\pmb \zeta$. This is given by
\begin{eqnarray}
&& dN_{\alpha\pmb \zeta \mathbf k}={\mathcal P}_{\alpha\pmb \zeta \mathbf k} \frac {\mathbf {k}^2 d|\mathbf {k}|}{\Phi^\alpha_{\mathbf k}} d\Omega_{\mathbf k},\\
&& {\mathcal P}_{\alpha\pmb \zeta \mathbf k}:= \sum_{\mu=1}^2 \sum_{\beta=0}^N \int
|\mathcal {A}_{\{ \alpha \pmb {\zeta} \mathbf{k}; \beta \mathbf {e}_{\beta\mu\mathbf {k'}} \mathbf{k'} \}}|^2
\frac {d^3 \mathbf k'}{\Phi^\beta_{\mathbf{k'}}}.
\end{eqnarray}
A direct computation gives
\begin{align}
&{\mathcal P}_{\alpha\pmb \zeta \mathbf k}=\frac 1{c^4} \sum_{\beta=0}^N \sum_{l=1}^N \sum_{s=1}^N \int
\widehat {\delta\chi}_{(l)} (\omega_{\alpha\mathbf {k}}+\omega_{\beta\mathbf {k'}}, \mathbf {k}+\mathbf {k'}) \cdot\cr
&\cdot\widehat {\delta\chi}_{(s)}^* (\omega_{\alpha\mathbf {k}}+\omega_{\beta\mathbf {k'}}, \mathbf {k}+\mathbf {k'}) \cdot \cr
& \cdot \frac {\omega_{\alpha\mathbf k}^2\omega_{\beta\mathbf {k'}}^2(\omega_{\alpha\mathbf k}\omega_{\beta\mathbf {k'}}+\omega^2_{0(l)})
(\omega_{\alpha\mathbf k}\omega_{\beta\mathbf {k'}}+\omega^2_{0(s)})}{(\omega^2_{0(l)}-\omega_{\alpha\mathbf k}^2)(\omega^2_{0(l)}-\omega_{\beta\mathbf {k'}}^2)
(\omega^2_{0(s)}-\omega_{\alpha\mathbf k}^2)(\omega^2_{0(s)}-\omega_{\beta\mathbf {k'}}^2)}\cdot\cr
&\cdot\left[ 1- \frac {(\mathbf {k'} \cdot \pmb \zeta_{\mathbf k})^2}{\mathbf {k'}^2} \right]\ \frac {d^3 \mathbf k'}{\Phi^\beta_{\mathbf{k'}}}.
\end{align}
If we are not interested in the polarization of the produced polaritons, we can sum over $\pmb \zeta$:
\begin{align}
&{\mathcal P}_{\alpha \mathbf k} = \frac 1{c^4} \sum_{\beta=0}^N \sum_{l=1}^N \sum_{s=1}^N \int
\widehat {\delta\chi}_{(l)} (\omega_{\alpha\mathbf {k}}+\omega_{\beta\mathbf {k'}}, \mathbf {k}+\mathbf {k'})\cdot\cr
&\cdot\widehat {\delta\chi}_{(s)}^* (\omega_{\alpha\mathbf {k}}+\omega_{\beta\mathbf {k'}}, \mathbf {k}+\mathbf {k'}) \cdot \cr
& \cdot \frac {\omega_{\alpha\mathbf k}^2\omega_{\beta\mathbf {k'}}^2(\omega_{\alpha\mathbf k}\omega_{\beta\mathbf {k'}}+\omega^2_{0(l)})
(\omega_{\alpha\mathbf k}\omega_{\beta\mathbf {k'}}+\omega^2_{0(s)})}{(\omega^2_{0(l)}-\omega_{\alpha\mathbf k}^2)(\omega^2_{0(l)}-\omega_{\beta\mathbf {k'}}^2)
(\omega^2_{0(s)}-\omega_{\alpha\mathbf k}^2)(\omega^2_{0(s)}-\omega_{\beta\mathbf {k'}}^2)}\cdot\cr
&\cdot\left[ 1+ \frac {(\mathbf {k'} \cdot \mathbf {k})^2}{\mathbf {k'}^2 \mathbf {k'}^2} \right]\ \frac {d^3 \mathbf k'}{\Phi^\beta_{\mathbf{k'}}}. \label{density}
\end{align}
Finally, by using the dispersion relation, for the number of polaritons with frequency $\omega_{\alpha\mathbf k} \leq \omega \leq \omega_{\alpha\mathbf k}+d\omega$
and direction $d\Omega_{\mathbf k}$ we get
\begin{eqnarray}
dN_{\alpha \mathbf k}={\mathcal P}_{\alpha \mathbf k} \frac {\omega_{\alpha\mathbf k}}{2c}
\frac {n_p(\omega_{\alpha\mathbf k})}{(2\pi)^3} d\omega d\Omega_{\mathbf k}.
\end{eqnarray}
Notice that in (\ref{density}) the measure factor avoids the poles in the denominators of the fraction in the second line, so that possible divergences depend
only on the first line. However, the denominators allow to individuate the main contributors to the integral.\\
An alternative interesting expression is the one predicting the number of photon pairs emitted in the cones $d\Omega_{\mathbf k}$, $d\Omega_{\mathbf k'}$,
with energies in the branches $\alpha$ and $\alpha'$, $dE_\alpha=\hbar d\omega_\alpha$, $dE_{\alpha'}=\hbar d\omega_{\alpha'}$:
\begin{align}
&dN_{\alpha \mathbf k \vec \zeta; \alpha' \mathbf k' \vec \zeta'}=\frac {\vec \zeta \cdot \vec \zeta'}{c^4} \sum_{l=1}^N \sum_{s=1}^N \left\{
\widehat {\delta\chi}_{(l)} (\omega_{\alpha\mathbf {k}}+\omega_{\alpha'\mathbf {k'}}, \mathbf {k}+\mathbf {k'}) \cdot\right.\cr
&\left. \cdot \widehat {\delta\chi}_{(s)}^* (\omega_{\alpha\mathbf {k}}+\omega_{\alpha'\mathbf {k'}}, \mathbf {k}+\mathbf {k'}) \cdot \right. \cr
& \left. \cdot \frac {\omega_{\alpha\mathbf k}^2\omega_{\alpha'\mathbf {k'}}^2(\omega_{\alpha\mathbf k}\omega_{\alpha'\mathbf {k'}}+\omega^2_{0(l)})
(\omega_{\alpha\mathbf k}\omega_{\alpha'\mathbf {k'}}+\omega^2_{0(s)})}{(\omega^2_{0(l)}-\omega_{\alpha\mathbf k}^2)(\omega^2_{0(l)}-\omega_{\alpha'\mathbf {k'}}^2)
(\omega^2_{0(s)}-\omega_{\alpha\mathbf k}^2)(\omega^2_{0(s)}-\omega_{\alpha'\mathbf {k'}}^2)} \right\} \cdot \cr
 & \cdot \frac {\omega_{\alpha\mathbf k}}{2c} \frac {n_p(\omega_{\alpha\mathbf k})}{(2\pi)^3}
\frac {\omega_{\alpha'\mathbf k'}}{2c} \frac {n_p(\omega_{\alpha'\mathbf k'})}{(2\pi)^3} d\omega_{\alpha} d\Omega_{\mathbf k} d\omega_{\alpha'} d\Omega_{\mathbf k'} .
\end{align}

\subsection{Reduced formulas for $N\leq 3$ resonances}

The general formulas we have obtained are of non-straight\-forward 
application for an arbitrary number $N$ of resonances. This is because the solutions $\omega_{\alpha \mathbf k}$ of the dispersion relation (\ref{dispersion-N})
can be obtained only numerically, being algebraic equations of order $N+1$ in $\omega^2$. 
However, in several applications one can physically put limits on the number of relevant resonances in given 
experimental situations, and, moreover, for $N\leq 3$ one can employ
the Cardano formulas. The case $N=3$ is indeed the interesting one when the dielectric material is fused silica. In this case the dispersion relation is described by
the Sellmeier relation
\begin{eqnarray}
\frac {c^2 \mathbf {k}^2}{\omega^2}=1+ \frac {a_1 \lambda^2}{\lambda^2-l_1^2} +\frac {a_2 \lambda^2}{\lambda^2-l_2^2} +\frac {a_3 \lambda^2}{\lambda^2-l_3^2},
\end{eqnarray}
with
\begin{eqnarray}
& a_1=0.906404498, & l_1=98.7685322\, {\mathrm \mu m}, \\
& a_2=0.473115591, & l_2=0.0129957170\, {\mathrm \mu m}, \\
& a_3=0.631038719, & l_3=4.12809220\cdot 10^{-3}\, {\mathrm \mu m}.
\end{eqnarray}
This corresponds to (\ref{dispersion-N}) with $N=3$,
\begin{align}
\omega_{0(l)}^2=\frac {4\pi^2 c^2}{l_l^2}, \qquad\ \chi_{0(l)}=a_l \omega_{0(l)}^2, \qquad l=1,2,3.
\end{align}
In physical situations, involved with 
photons whose frequency is well below the lowest resonance pole of the dispersion relation, 
the relevant contributions are associated only with the lowest branch of the dispersion relation. 
Typically, this happens in diamond when the frequencies  
of the photons involved in the physical situation at hand 
are well below the lowest resonance pole. 
In this case, the number of emitted pairs assumes the simpler form
\begin{align}
&dN_{\alpha \mathbf k \vec \zeta; \alpha' \mathbf k' \vec \zeta'}\!\!\!=\!\!\! \frac {\vec \zeta \cdot \vec \zeta'}{4(2\pi c)^6}
|\widehat {\delta\chi} (\omega_{\alpha\mathbf {k}}+\omega_{\alpha'\mathbf {k'}}, \mathbf {k}+\mathbf {k'})|^2\cdot\cr
&\cdot\frac {\omega_{\alpha\mathbf k}^2\omega_{\alpha'\mathbf {k'}}^2(\omega_{\alpha\mathbf k}\omega_{\alpha'\mathbf {k'}}+\omega^2_{0})^2
}{(\omega^2_{0}-\omega_{\alpha\mathbf k}^2)^2(\omega^2_{0}-\omega_{\alpha'\mathbf {k'}}^2)^2}\cdot \cr
&\cdot{\omega_{\alpha\mathbf k}} {n(\omega_{\alpha\mathbf k})}
 {\omega_{\alpha'\mathbf k'}}  {n(\omega_{\alpha'\mathbf k'})} d\omega_{\alpha} d\Omega_{\mathbf k} d\omega_{\alpha'} d\Omega_{\mathbf k'},
\label{dn-diamond}
\end{align}
where now $\alpha, \alpha'$ assume the values $\pm$ and $\omega_\pm^2$ coincides with (\ref{d-branches}).

\section{Photon pair creation by an uniformly travelling dielectric perturbation}

As an example, let us consider the case of a refractive index perturbation moving propagating along the $z$ direction with constant velocity $v$. 
The model we further explore herein was introduced in \cite{belgio-prl}, and it is based on the idea that a travelling dielectric perturbation, 
which is induced by an intense laser pulse which passes through a dielectric medium, is able to generate photon pairs. The original model 
involved a nondispersive medium, and a phenomenological approach to the electromagnetic field quantization. We improve that model, by showing that, 
in a framework including automatically optical dispersion, rooted into microscopical characteristics of the matter fields, photon pair production is 
ensured. We choose to simulate a perturbation of the refractive index by means of a perturbation in the dielectric susceptibility $\chi$, and in 
particular we assume that it is  of the form
\begin{eqnarray}
\delta\chi(t,\rho, z, \phi) =\delta \chi_{0} e^{-\frac {\rho^2}{2\sigma^2_\rho}} e^{-\frac {(z-z(t))^2}{2\sigma^2_z}},
\end{eqnarray}
where $z(t)$ is an arbitrary motion.  This Gaussian form can be easily justified in the nondispersive case, where $n_0 (n_0+2\delta n) \sim 
1+\chi_0+\delta \chi$, i.e. $2 n_0\delta n \sim \delta \chi$, where $n_0$ is the unperturbed (constant) refractive index. By means of the well known Weber formula
\begin{eqnarray}
\int_{0}^\infty \rho e^{-\frac {\rho^2}{2\sigma_\rho^2}} J_0 (\rho k_\rho) d\rho =\sigma_\rho^2 e^{-\frac {\sigma_\rho^2 k_\rho^2}2}
\end{eqnarray}
it is easy to compute the Fourier transform of the perturbation:
\begin{align}
\widehat {\delta\chi}(\omega, k_\rho, k_z, k_\phi)&=(2\pi)^{\frac 32} \delta \chi_{0} \sigma_z \sigma_\rho^2
e^{-\frac {\sigma_z^2 k^2_z}2-\frac {\sigma_\rho^2 k^2_\rho}2}\cdot\cr
&\cdot \int_{\Delta T} e^{i(\omega_{\mathbf k} t-k_z z(t))} dt,
\end{align}
where the $\Delta t\equiv [-T,T]$ is the time duration of the perturbation inside the Kerr dielectric matter. The characterizing information
is thus contained in the term
\begin{eqnarray}
f_T(\omega_{\mathbf k}):=\int_{-T}^T e^{i(\omega_{\mathbf k} t-k_z z(t))} dt.
\end{eqnarray}
We now suppose that the perturbation is moving with constant velocity $v$.
In this case
\begin{eqnarray}
f_T(\omega_{\mathbf k})=2 \frac {\sin [(\omega_{\mathbf k}-k_z v)T]}{\omega_{\mathbf k}-k_zv},
\end{eqnarray}
so that, for large $T$
\begin{eqnarray}
|f_T(\omega_{\mathbf k})|^2\simeq T\pi \delta (\omega_{\mathbf k}-k_zv),
\end{eqnarray}
and the number of pairs for unit time emitted with momenta $\mathbf k$ and $\mathbf k'$ in the angles
$d\Omega_{\mathbf k}$ and $d\Omega_{\mathbf k'}$, is
\begin{align}
\frac {dN}{2T} &= \frac {\vec \zeta \cdot \vec \zeta'}{16\pi^2 c^6} (\delta \chi_{0})^2
\frac {\omega_{\alpha\mathbf k}^2\omega_{\alpha'\mathbf {k'}}^2(\omega_{\alpha\mathbf k}\omega_{\alpha'\mathbf {k'}}+\omega^2_{0})^2
}{(\omega^2_{0}-\omega_{\alpha\mathbf k}^2)^2(\omega^2_{0}-\omega_{\alpha'\mathbf {k'}}^2)^2} \cdot \cr
&\cdot \sigma_z^2 \sigma_\rho^4
e^{-\sigma_z^2(k_z+k'_z)^2} e^{-\sigma_\rho^2(k_\rho+k'_\rho)^2} \cdot \cr
& \cdot \delta (\omega_{\alpha\mathbf k}+\omega_{\alpha'\mathbf k'} -vk_z-vk'_z) \cdot \cr
&\cdot \omega_{\alpha\mathbf k} n_p(\omega_{\alpha\mathbf k}) \omega_{\alpha'\mathbf k'} n_p(\omega_{\alpha'\mathbf k'})
\delta\omega_{\alpha} d\Omega_{\mathbf k} \delta\omega_{\alpha'} d\Omega_{\mathbf k'}. \label{formulazza}
\end{align}
This expression can be used to simplify the analysis and confirm the results obtained in \cite{DallaPiazza:2012aj}.
From (\ref{formulazza}) we can also more readily get further information about the emitted spectrum. For example, from the gaussian terms
we see that a large pulse, with a large $\sigma_\rho$, gives rise to conservation of the transversal components of the momentum
$k_\rho+k'_\rho\approx 0$. However, the Dirac $\delta$ function does not allows for conservation of the $z$ component of the momentum
(unless $\omega=\omega'\approx 0$). Thus, to have a significant emission of photon one should produce a short pulse with a small $\sigma_z$
parameter.
As pointed out in \cite{belgio-prl} and in \cite{DallaPiazza:2012aj}, the perturbative analysis indicates that the pairs production  occurs only if $v>c/n(\omega)$, i.e. only if the perturbation of the refractive index is superluminal, and the number of emitted particles increases with $v$.
Furthermore, we observe that the argument of the delta function in equation \eqref{formulazza} is the same as in \cite{belgio-prl} (cf. equations (7) and (10)) and in \cite{DallaPiazza:2012aj} (cf. equations (26), (29) and (30)). 
We recall here its interpretation and its interesting physical meaning. The support of the delta distribution gives a constraint on the state of the emitted particles in the pair. In the non dispersive case, employing the relation $k=\frac{\omega}{c}n_0$, we can rewrite the argument of the $\delta$ as $\left(k_z-\frac{c}{vn_0}k \right)+\left(k_z'-\frac{c}{vn_0}k' \right)=0$. This equation indicates that if $k_z/k>c/(vn_0)$, the momentum of the second photon must satisfy  $k_z'/k'<c/(vn_0)$. Thus, we obtain a cone structure for the distribution of the momenta of the emitted particles in a pair: one photon is emitted inside the Cerenkov cone, $\theta_0=\arccos(cv/n_0)$, and the other is emitted outside the cone. Due to the dependence of the refractive index on the frequency of the radiation, the dispersive case is more involved and in general one can not identify distinct cones of emission as in the non dispersive case. 
The constraint given by the $\delta$ distribution now is $\left(k_z-\frac{c}{vn(\omega)}k \right)+\left(k_z'-\frac{c}{vn(\omega')}k' \right)=0$. As before, this equation implies that whenever $k_z/k>c/[vn(\omega)]$, the momentum of the second photon must satisfy $k_z'/k'<c/[vn(\omega')]$. From the two conditions we obtain for the angle of emission, $\theta<\arccos\{c/[vn(\omega)]\}$ and $\theta'>\arccos\{c/[vn(\omega')]\}$. Thus, if $\arccos\{c/[vn(\omega)]\}>\arccos\{c/[vn(\omega')]\}$, the two cones overlap; there is a gap between them if instead $\arccos\{c/[vn(\omega)]\}<\arccos\{c/[vn(\omega')]\}$. These considerations show that in the first case there is a region in which both photons can be emitted, instead in the second case there is a region in which no photon can be emitted. The presence of these two behaviors, depending on the frequencies of the emitted particles, makes the dispersive case interesting and substantially different from a non dispersive model. Obviously,  the non dispersive case can be seen as a limit case of the dispersive one.
Moreover, compared with the approach adopted in \cite{DallaPiazza:2012aj}, this one has the advantage to be available at any perturbative order.

\section{Case $\delta \chi (t)$}
\label{t-dep}

We can also consider a perturbation which depends only on time. 
This case can be of noticeable physical interest, in view of the possibility to induce (locally) 
purely time-dependent perturbations in optical systems. For simplicity, we focus explicitly 
on the case of a diamond-like dielectric. Extensions to more general cases are indeed straightforward. First of all, we take into account that the Fourier transform of the perturbation 
is non-trivial only in $t$, of course. As a consequence, we get
\beq
\widehat {\delta\chi}(\omega, \mathbf k)=\delta^3 (\mathbf k) \widehat {\delta\chi}(\omega).
\eeq
Then, in (\ref{dn-diamond}) we get that the square modulus of $\widehat {\delta\chi}$ is 
\beq
|\widehat {\delta\chi}(\omega_{\alpha\mathbf k}+\omega_{\alpha'\mathbf k'})|^2 
\delta^3 (\mathbf k+\mathbf k') \delta^3  (0)
\label{dn-time}
\eeq
($\delta^3  (0)$ is to be replaced by a volume factor, as usual). 
As it is evident, pairs are produced back-to-back in this situation. See below for a more 
general case, accounting for finite-size effects.\\
Interesting examples of time dependence are the following (with $\eta\ll 1$ constant): a) a Gaussian dependence in time 
\beq
\delta \chi (t) = \eta \exp (-a t^2), \quad \quad a>0,
\eeq
so that 
\beq
\widehat {\delta\chi}(\omega)=\eta \frac{\pi}{\sqrt{ \omega}} \exp \left(-\frac{\omega^2}{4 a}\right)
\eeq
(it simulates a perturbation which is peaked around $t\sim 0$ and is quite soon zero for 
$t\not \sim 0$). b) Another interesting perturbation profile 
is a step-like perturbation, which allows to deal with the case of a rapidly rising perturbation 
and to calculate the number of produced pairs in the raising phase. For example, we can 
adopt the profile
\beq
\delta \chi (t) =\eta (1+\tanh (a t)),
\eeq
which provides
\beq
\widehat {\delta\chi}(\omega)=\eta \left[i \frac{\pi}{a} \frac{1}{\sinh \left( \frac{\pi \omega}{2 a}\right)} + {2 \pi} \delta (\omega) \right]. 
\eeq
c) As a further interesting perturbation, we could consider a periodic perturbation:
\beq
\delta \chi (t)= \eta (1+\sin (a t)),
\eeq
whose Fourier transform is  
\beq
\eta \left[{2 \pi} \delta (\omega)+ i {\pi} \delta (\omega - a)
-i {\pi} \delta (\omega +a) \right].
\eeq
It is evident that photon production, in this specific case, happens only  at resonances: $\omega=\pm a$.
It is also interesting to note that our picture can be easily generalized to the case of a 
perturbation which has finite spatial support (instead of being extended to all the space). 
The only difference consists in the fact that pair-emission is not strictly confined to 
be back-to-back, due to finite-size effects. Indeed, if we assume that the perturbation is 
\beq
\delta \chi (t) \gamma (\mathbf x),
\eeq
where $\gamma$ has e.g. compact spatial support, we obtain a Fourier transform 
$\widehat{\gamma}(\mathbf k)$ which is related to a pair-emission non-strictly 
back-to-back, due to finite-size effects. Indeed, in (\ref{dn-diamond})  
the factor $|\widehat{\gamma}(\mathbf k + \mathbf k')|^2$ replaces 
$\delta^3 (\mathbf k+\mathbf k') \delta^3 (0)$ appearing in (\ref{dn-time}).

\section{Conclusions}

We have explored, in a perturbative framework, a covariant generalization of the Hopfield model aimed to modelize, in a less 
phenomenological way, photon pair creation phenomena associated with dielectric media with spacetime dependent 
dielectric constant. This dependence can be realized in different ways, and we can refer both to Kerr effect in nonlinear dielectric 
media, and to sonoluminescence. The advantage of the model, with respect to the ones existing in literature, is that 
optical dispersion, which necessarily plays a role in any physical settings, is automatically taken into account, as well as 
covariance of the results. In this sense, even if with the limitation that only dielectric properties are taken into account by the 
present model, our results generalize 
the ones in \cite{SPS} to the case of dispersive media. The general expressions we have found can be applied to
several situations where dispersion becomes relevant.\\
Moreover, Lorentz covariance could be employed to reexpress all results in any inertial frame, as the comoving one 
in the example we have provided. Further interesting applications will be
presented elsewhere \cite{Helix}.

\section*{Acknowledgement}
We thank D. Faccio for discussions and for suggesting us the picture of purely time dependent 
profiles of section (\ref{t-dep}).


\end{document}